\documentclass[aps,preprint]{revtex4-1}
\usepackage{amsmath}
\usepackage{graphicx}
\usepackage{amssymb}

\usepackage{gt17}
\usepackage{color} 
\usepackage{graphicx}% Include files
\usepackage{bm}% bold math
\begin{document}
%%%%%%%%%%%%%%%%%%%%%%%%%%
%%%%%%%%%%%%%%%%%%%%%%%%%%%%%%%%%%%%%
\title{
Collective coordinate study of spin wave emission from dynamic domain wall}
\author{Gen Tatara} 
%\email{gen.tatara@riken.jp}
\affiliation{RIKEN Center for Emergent Matter Science (CEMS)
%and RIKEN Cluster for Pioneering Research (CPR), 
2-1 Hirosawa, Wako, Saitama, 351-0198 Japan}
\author{Rubén M. Otxoa de Zuazola}
\affiliation{Hitachi Cambridge Laboratory, J. J. Thomson Avenue, CB3 OHE, Cambridge, United Kingdom}
\affiliation{Donostia International Physics Center,  20018 San Sebasti\'an, Spain}
\date{\today}

\begin{abstract}
We study theoretically the spin wave emission from a moving domain wall in a ferromagnet. 
Introducing a deformation mode describing a modulation of the wall thickness in the collective coordinate description, we show that thickness variation couples to the spin wave linearly and induces spin wave emission.
The dominant emitted spin wave turns out to be polarized in the out-of wall plane ($\phi$)-direction.
The emission contributes to the Gilbert damping parameter proportional to $\hbar\omega_\phi/K$, the ratio of the angular frequency $\omega_\phi$ of $\phi$ and the easy-axis anisotropy energy $K$.
\end{abstract}  
%\pacs{78.20.-e, 71.70.Ej, 85.75.-d, 81.05.Xj}

\maketitle
%%%%%%%%%%%%%%
\section{Introduction}
%%%%%%%%%%%%%%%%%%%%%%%%%%%%%%%%%%%%%%%%%%%%%%%%%%
Spin wave (magnon) is an excitation playing essential roles in the transport phenomena in magnets, and its control, magnonics, is a hot recent issue. 
Besides application interest for devices, behaviours of spin waves have been drawing interests from fundamental science view points.  
Many theoretical studies have been carried out on generation of spin waves by dynamic magnetic objects such as a domain wall \cite{Bouzidi90,LeMaho09,Wang12,Wang14,Whitehead17,KimSW18}. 
The subject is highly nontrivial because the wall is a soliton, which is stable in the absence of perturbation, meaning that it couples to fluctuations, spin waves, only weakly in the ideal case, while in reality, various perturbations and dynamics leads to strong emission of spin waves. 
There are several processes that lead to the emission, and 
it is not obvious which is the dominant process and how large is the dissipation caused by the emission.

The low energy behavior of a domain wall in a ferromagnet is described in terms of collective coordinates, its center of mass position $X$ and angle of the wall plane, $\phi_0$ \cite{Slonczewski72}. 
In the absence of a pinning potential,  a displacement of the wall costs no energy owing to the translational invariance, and it is thus natural to regard $X$ as a dynamic variable $X(t)$.
This is in fact justified mathematically;  $X(t)$ is a collection of spin waves that corresponds to the translational motion of the wall \cite{Rajaraman82,TKS_PR08}. 
It turns out that the canonical momentum of the ferromagnetic domain wall is the angle $\phi_0$. 
This is because the translational motion of collective spins requires a perpendicular spin polarization, i.e., a tilting of the wall plane. 
Mathematically this is a direct consequence of the spin algebra, and is straightforwardly derived based on the equation of motion for spin (Landau-Lifshitz(-Gilbert) equation) \cite{Slonczewski72} or on the Lagrangian formalism \cite{TK04}.
In the absence of hard-axis anisotropy energy, $\phi_0$ is also a zero mode. 
As zero modes,  $X(t)$ and $\phi_0$ do not have linear coupling to the fluctuation, spin wave, and thus emission of spin wave does not occur to the lowest order.
In this case, the second-order interactions to the spin wave give rise to the dominant effect.  
In Ref. \cite{Bouzidi90}, the coupled equations of motion for the wall and spin wave modes were solved classically and demonstrated that a damping indeed arises from the quadratic interaction. 
In the case of a strong hard-axis anisotropy, the plane of the wall is constrained near the easy-plane, $\phi_0$ is frozen, resulting in a single variable system described solely by $X(t)$ \cite{Braun96,TKS_PR08}.  
The spin wave coupling and dissipation in this limit was discussed in Ref. \cite{Braun96}.

In real materials, hard-axis anisotropy and pinning potential exist, and  $X(t)$ and $\phi_0$ are not rigorously zero modes.
In other words, wall dynamics induces a deformation and emission of spin wave is possible due to linear couplings. 
It was argued in Ref. \cite{LeMaho09} that there emerges a linear coupling when the wall driven by a spin-transfer torque has a velocity $\dot{X}$ different from the steady velocity determined by the spin-transfer torque, and the damping due to spin wave emission was discussed. 
Numerical analysis of Ref. \cite{Wang12} revealed that spin wave emission occurs by the modulation of the wall thickness during the dynamics. 
The coupling to the wall velocity and second order in the spin wave was studied analytically in detail and dissipation was estimated in Ref. \cite{KimSW18}. The energy dissipation proportional to the second-order in the wall velocity was found.

In this paper, we study the spin wave emission extending conventional collective coordinate representation of the wall \cite{TK04}. 
As the domain wall is a soliton, there is no linear coupling of its center of mass motion to the spin wave field if   deformation is ignored. 
We thus introduce a deformation mode of the wall, a change of the thickness $\lambda$.
This is a natural variable in the presence of the hard-axis anisotropy energy, as the thickness depends on the angle $\phi_0$ as pointed out in Refs. \cite{Schryer74,Thiaville04}.    
Following the prescription of spin wave expansion \cite{TKS_PR08}, we derive the Lagrangian for the three collective coordinates, the center of mass position $X(t)$, the angle of the wall plane $\phi_0(t)$ and thickness 
$\lambda(t)$, including the spin waves to the second order. 
It turns out that $X$ and $\phi_0$ and their time-derivatives do not have linear coupling to the spin wave, while $\dot{\lambda}$ does. This result is natural as $X$ and $\phi_0$ are (quasi) zero modes, and consistent with numerical observation \cite{Wang12}. 
It is shown that the emitted spin wave is highly polarized; The dominant emission is the fluctuation of angle $\phi$, while that of $\theta$ is smaller by the order of the Gilbert damping parameter $\alpha$.
The forward emission of wavelength $\lambda^* \propto v_{\rm w}^{-1}$, where $v_{\rm w}$ is the domain wall velocity,  is dominant. 
The modulation of $\lambda$ is induced by the dynamics of $\phi_0$, and  
the contribution to the Gilbert damping parameter due to the spin wave emission from this process  
is estimated from the energy dissipation rate.
It was found to be 
of the order of $\alpha_{\rm sw}^\phi \simeq\frac{\lambda}{a}\frac{\hbar\omega_\phi}{K}$, where $\omega_\phi$ is the angular frequency of the modulation of $\phi_0$, $K$ is the easy-axis anisotropy energy and  $a$ is the lattice constant. 
This damping parameter contribution becomes very strong of the order of unity if $\hbar \omega_\phi$ is comparable to the spin wave gap, $K$, as deformation of the wall becomes significant in this regime. 

%%%%%%%%%%%%%%%%%%%%%%%%%%%%%%%%%%%%%%%%%%%%%%%%%%
\section{Collective coordinates for a domain wall}
%%%%%%%%%%%%%%%%%%%%%%%%%%%%%%%%%%%%
We consider a one-dimensional ferromagnet along the $x$-axis with easy and hard axis anisotropy energy along the $z$ and $y$ axis, respectively.
The Lagrangian in terms of polar coordinates $(\theta,\phi)$ of spin is
\begin{align}
 L=L_{\rm B}-H_{S} 
\end{align}
where
\begin{align}
L_{\rm B} & = \frac{\hbar S}{a} \int dx\dot{\phi}(\cos\theta-1) \nnr
H_{S} &= \frac{S^2}{2a}\int dx\lt[J[(\nabla \theta)^2+\sin^2\theta(\nabla \phi)^2] +K\sin^2\theta(1+\kappa\sin^2\phi)\rt]
\end{align}
are the kinetic term of the spin (spin Berry phase term) and the Hamiltonian, respectively, $J>0$, $K>0$ and $\kappa K\geq0$ being the exchange, easy-axis anisotropy and hard axis anisotropy energies, respectively, $a$ being the lattice constant.
A static domain wall solution of this system is 
\begin{align}
 \cos\theta &= \tanh\frac{x-X}{\lambda_0} , \phi=0
\end{align}
where $\lambda_0\equiv \sqrt{J/K}$ is the wall thickness at rest.
The dynamics of the wall is described by allowing the wall position $X$ and $\phi$ as dynamic variables. This corresponds to treat the energy zero mode of spin waves (zero mode) describing a translational motion and its conjugate variable $\phi$ as collective coordinates \cite{TKS_PR08}. This treatment is rigorous in the absence of pinning and hard-axis anisotropy but is an approximation otherwise.
Most previous studies considered a rigid wall, where the wall thickness is a constant $\lambda_0$. 
Here we  treat the wall thickness as a dynamic variable $\lambda(t)$ to include a deformation and study the spin-wave emission. 
This treatment was applied in Ref. \cite{Thiaville04}, but only static solution of $\lambda$ was discussed.

\begin{figure}
 \begin{minipage}{0.4\hsize}\centering
 \includegraphics[width=0.8\hsize]{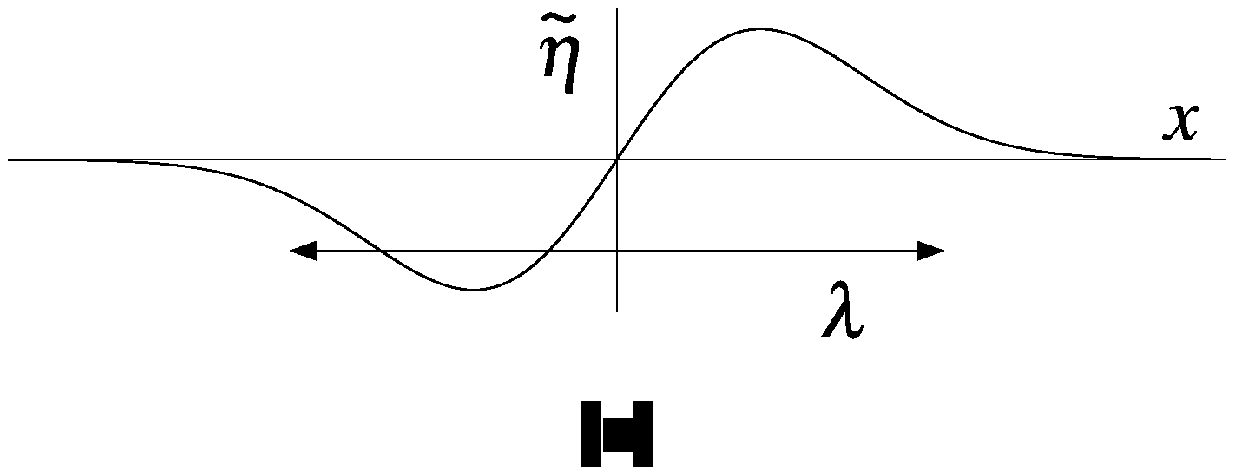}  
 \end{minipage}
 \begin{minipage}{0.4\hsize}\centering
 \includegraphics[width=0.8\hsize]{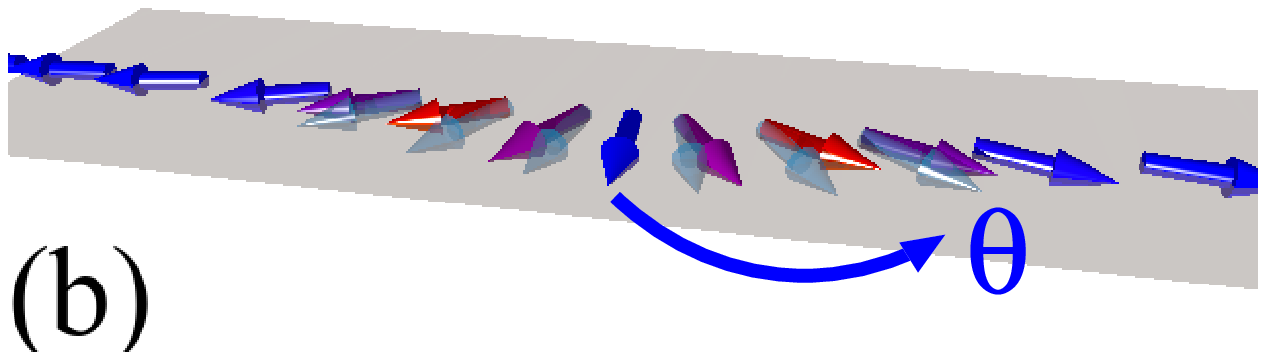}
 \includegraphics[width=0.8\hsize]{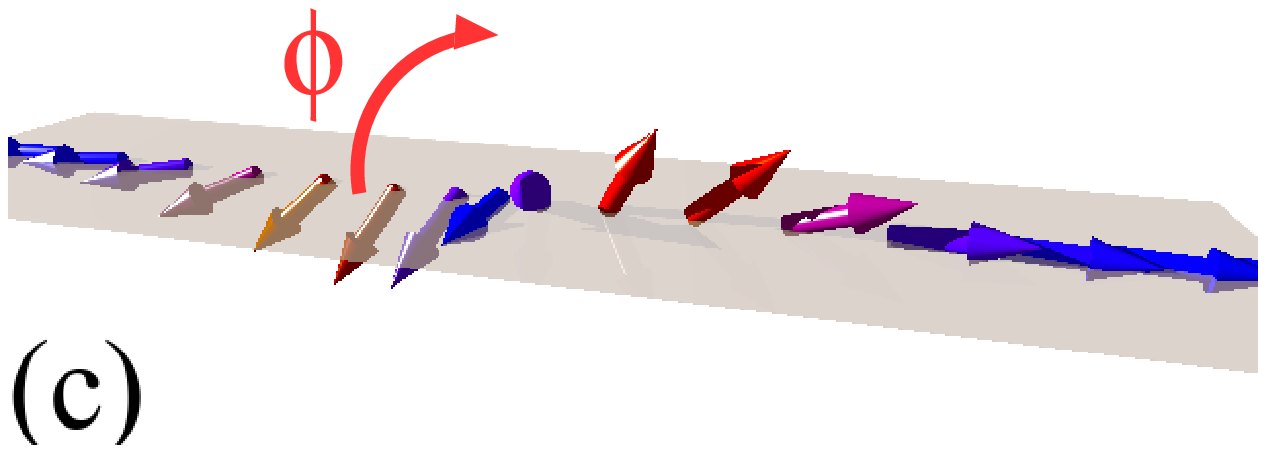}
 \end{minipage}
 \caption{ Fluctuation corresponding to the real and imaginary part of the spin wave variable $\tilde{\eta}=\tilde{\eta}_{\rm R}+i \tilde{\eta}_{\rm I}$. 
 (a): The profile of $\tilde{\eta}$ antisymmetric with respect to the wall center, which turns out to be dominant excitation.
 (b): The real part $\tilde{\eta}_{\rm R}$ describes the deformation within the wall plane, i.e., modulation of $\theta$, while the imaginary part $\tilde{\eta}_{\rm I}$ describes the out-of plane ($\phi$) fluctuation as shown in (c).  
 Transparent arrows denotes the equilibrium spin configuration.
 \label{FIGSW}} 
\end{figure}

As demonstrated in Ref. \cite{TKS_PR08}, the spin wave around a domain wall in ferromagnet is conveniently represented using
\begin{align}
 \xi=e^{-u(x,t)+i\phi_0(t)+\eta(x-X(t),t)} \label{xidef}
\end{align}
where $\phi_0(t)$ is the angle of the wall, 
\begin{align}
 u(x,t)&=\frac{x-X(t)}{\lambda(t)}
\end{align}
and $\eta(x-X,t)$ describes the spin-wave viewed in the moving frame.  
As it is obvious from the definition, the real and imaginary part of $\eta$ describe the fluctuation of $\theta$ and $\phi$, respectively. 
The fluctuations antisymmetric with respect to the wall center, shown in Fig. \ref{FIGSW}, turns out to be dominant.
The $\xi$-representation of the polar angles are 
\begin{align}
 \cos\theta &= \frac{1-|\xi|^2}{1+|\xi|^2}, &  \sin\theta \sin\phi = -i\frac{\xi-\overline{\xi}}{1+|\xi|^2}. 
\label{thetaxi}
\end{align}

%%%%%%%%%%%%%%%%%%%%%%%%%%
%%%%%%%%%%%%%%%%%%%%%%%%%%%%%%%%%%%%%%
%%%%%%%%%%%%%%%%%%%%%%%%%%%%%%%

%%%%%%%%%%%%%%%%%%%%%%%%%%%%%%%%%%%%%%%%%%%%%%%%%%
\subsection{Domain wall dynamic variables \label{SECdynamicvariable}}
%%%%%%%%%%%%%%%%%%%%%%%%%%%%%%%%%%%%
We first study what spin-wave mode the new variable $\lambda(t)$ couples to, by investigating  the 'kinetic' term of the spin Lagrangian, $L_{\rm B}$, which is written as 
\begin{align}
 L_{\rm B} &= \frac{2i\hbar S \lambda}{a} \int du \frac{\Im[\overline{\xi}\dot{\xi}]}{1+|\xi|^2}. 
\end{align}
Using Eq. (\ref{thetaxi}) and 
\begin{align}
 \partial_t u &= -\frac{1}{\lambda}\lt(\dot{X}+u\dot{\lambda}\rt), &
 \partial_t \xi &= \lt(\frac{1}{\lambda}\lt(\dot{X}+u\dot{\lambda}\rt)+i\dot{\phi_0}+(\partial_t-\dot{X}\nabla_x)\eta\rt)\xi, 
\end{align}
we have
\begin{align}
2i\Im[ \overline{\xi}\dot{\xi}] &=2i(\dot{\eta_{\rm I}}+\dot{\phi_0}-\dot{X}\nabla_x\eta_{\rm I})|\xi|^2
\end{align}
where $\eta_{\rm i}\equiv \Im[\eta]$.
The kinetic term is expanded to the second order in the spin wave as (using integral by parts) 
\begin{align}
 L_{\rm B} &= \frac{2\hbar S }{a} [ {\phi_0}\dot{X}+ \varphi \dot{\lambda} ] +L_{\rm B}^{(2)}
 \label{LB1}
\end{align}
where 
\begin{align}
 \varphi\equiv \int du \frac{u}{\cosh u}\tilde{\eta_{\rm I}},
 \label{varphidef}
\end{align}
represents an asymmetric configuration of $\tilde{\eta_{\rm I}}$ 
and
\begin{align}
L_{\rm B}^{(2)} &\equiv 
\frac{2\hbar S \lambda}{a} \int du \lt[\tilde{\eta_{\rm R}} \vvec{\partial_t} \tilde{\eta_{\rm I}} 
-\dot{X} \tilde{\eta_{\rm R}} \vvec{\nabla_x} \tilde{\eta_{\rm I}} 
-\frac{2}{\lambda}\tanh u \lt( 2{\dot{X}}+ u{\dot{\lambda}}\rt)\tilde{\eta_{\rm R}} \tilde{\eta_{\rm I}}
  \rt],
  \label{L2result}
\end{align}
 where $\tilde{\eta}\equiv \eta/(2\cosh u)$. 
 
\begin{figure}
 \includegraphics[width=0.5\hsize]{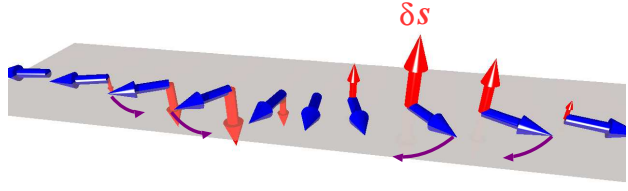}
 \caption{ Schematic figure showing the effect of asymmetric perpendicular spin polarization $\delta s$ due to the spin wave mode 
 $\varphi$. The asymmetric torque (curved arrows) induced by asymmetric $\delta s$ rotates the spins within the wall plane, resulting in a compression of the wall, i.e., to $\dot{\lambda}$.
 \label{FIGcompress}} 
\end{figure}

When deriving Eq. (\ref{LB1}), the orthogonality of fluctuation and the zero-mode,
\begin{align}
\int du \frac{\tilde{\eta}}{\cosh u}=0, \label{orthogonality}
\end{align}
 was used. 
Equation (\ref{LB1}) indicates that $\varphi$ is the canonical momentum of $\lambda$. 
In fact, it represents the asymmetric deformation of angle $\phi$, as the imaginary part of the spin wave, $\tilde{\eta_{\rm I}}$, corresponds to fluctuation of $\phi$ as seen in the definition, Eq. (\ref{xidef}). 
Such an asymmetric configuration of $\phi$ exerts a torque that induces a compression or expansion of the domain wall (Fig. \ref{FIGcompress}), and this is why $\varphi$ and $\lambda$ are conjugate to each other. 
The coupling $\varphi \dot{\lambda} $ describes the spin wave emission when thickness changes, as we shall argue later. 
The second term proportional to $\dot{X}$ in the bracket in Eq. (\ref{L2result}) represents a magnon current induced in the moving frame (Doppler shift).

%%%%%%%%%%%%%%%%%%%%%%%%%%
%%%%%%%%%%%%%%%%%%%%%%%%%%%%%%%%%%%%%%
%%%%%%%%%%%%%%%%%%%%%%%%%%%%%%%

The Hamiltonian of the system is similarly written in terms of spin wave variables to the second order as 
\begin{align}
H_{S} &=
\frac{K S^2 \lambda}{a}\lt[ \lt(\frac{\lambda_0}{\lambda}\rt)^2+1+\kappa \sin^2\phi_0\rt] 
+2\frac{K S^2 \lambda}{a}\int du \frac{\tanh u}{\cosh u} \tilde{\eta_{\rm R}} 
 \lt[ -\lt(\frac{\lambda_0}{\lambda}\rt)^2+1+\kappa \sin^2\phi_0\rt] +H_{S}^{(2)},
 \label{HSresult}
\end{align}
where %$\lambda_0\equiv \sqrt{{J}/{K}}$ is the domain wall thickness at the equilibrium and 
\begin{align}
H_{S}^{(2)} &\equiv 
2\frac{K S^2 \lambda}{a}\int du \biggl[ \lambda_0^2[ (\nabla \tilde{\eta_{\rm R}})^2 + (\nabla \tilde{\eta_{\rm I}})^2 ] \nnr
&+\tilde{\eta_{\rm R}}^2 \lt[-\frac{\lambda_0^2}{\lambda^2} \lt(1-\frac{1}{\cosh^2 u}\rt)
+\lt(2-\frac{3}{\cosh^2 u}\rt)(1+\kappa\sin^2\phi_0) \rt]
+\tilde{\eta_{\rm I}}^2 \lt[\frac{\lambda_0^2}{\lambda^2} \lt(1-\frac{2}{\cosh^2 u}\rt) + \kappa \cos 2\phi_0\rt]
\nnr
& +2\kappa \tilde{\eta_{\rm R}}\tilde{\eta_{\rm I}} \tanh u \sin 2\phi_0 \biggr]
\label{H2result}
\end{align}
In the case of small $\kappa$ and $\lambda\simeq \lambda_0$, the spin waves are described by a simple Hamiltonian as
\begin{align}
H_{\rm sw} &\equiv 
2\frac{K S^2 \lambda}{a}\int du \biggl[ \lambda^2[ (\nabla \tilde{\eta_{\rm R}})^2 + (\nabla \tilde{\eta_{\rm I}})^2 ] 
+(\tilde{\eta_{\rm R}}^2 +\tilde{\eta_{\rm I}}^2) \lt(1-\frac{2}{\cosh^2 u}\rt)\biggr]
+H_{\rm D},\label{Hsw}
\end{align}
where 
\begin{align}
H_{\rm D} &\equiv 
\frac{2\hbar S \lambda}{a} \dot{X} \int du  \tilde{\eta_{\rm R}} \vvec{\nabla_x} \tilde{\eta_{\rm I}} ,
\end{align}
is the Doppler shift term. For a constant wall velocity $\dot{X}$, it simply shifts the wave vector of the spin wave. 
Without the Doppler shift, the eigenfunction of this Hamiltonian (\ref{Hsw}) is labeled by a wave vector $k$ as 
\begin{align}
 \phi_k(u)&=\frac{1}{\sqrt{2\pi\tilde{\omega}_k}}(-ik\lambda+\tanh u)e^{ik\lambda u}, 
 \label{SWeigenfunc}
\end{align}
where 
\begin{align}
 \tilde{\omega}_k\equiv 1+(k\lambda)^2
\end{align}
is the dimensionless energy of spin wave.

Dissipation function is 
\begin{align}
 W &= \frac{\alpha \hbar S}{2a}\int dx (\dot{\theta}^2+\sin^2\theta \dot{\phi}^2) \nnr
 &= \frac{\hbar S \lambda}{2a} \lt[\alpha \lt(\frac{\dot{X}}{\lambda}\rt)^2+\alpha \dot{\phi_0}^2
  +\alpha_\lambda \lt(\frac{\dot{\lambda}}{\lambda}\rt)^2 \rt],\label{Wresult}
\end{align}
where $\alpha$ is the Gilbert damping parameter and 
$\alpha_\lambda\equiv \alpha\int du \frac{u^2}{\cosh ^2 u}=\frac{\pi^2}{12}\alpha$.

As driving mechanisms of a domain wall, we consider a magnetic field and current-induced torque (spin-transfer torque) \cite{Berger86,TKS_PR08,TataraReview19}. 
A magnetic field applied along the negative easy axis is represented by the Hamiltonian ($\gamma=e/m$ is the gyromagnetic ratio) 
\begin{align}
 H_B &= \frac{\hbar S \gamma}{a}B_z \int dx \cos\theta.
\end{align}
Using Eqs. (\ref{thetaxi})(\ref{orthogonality}), we obtain 
\begin{align}
 H_B &= -\frac{2\hbar S \gamma}{a}B_z \lt( X+ \lambda \int du \tanh u \;  \tilde{\eta}_{\rm R}^2 \rt). 
 \label{HBsw}
\end{align}
(The first term is derived evaluating a diverging integral $\int dx \frac{1}{1+e^{2u(x)}}$ carefully introducing the system size $L$ as  $\int^{L/2}_{-L/2} dx \frac{1}{1+e^{2u(x)}}$ and dropping a constant.)
The magnetic field therefore exerts a force $\frac{2\hbar S \gamma}{a}B_z$ on the domain wall.

The spin-transfer effect induced by injecting spin-polarized electric current is represented by a Hamiltonian having the same structure as the spin Berry's phase term $L_{\rm B}$ \cite{TKS_PR08,TataraReview19}
\begin{align}
 H_{\rm STT} &= -\frac{\hbar S}{a}v_{\rm st} \int dx \cos\theta(\nabla_x \phi),
\end{align}
where $v_{\rm st}\equiv \frac{aP}{2eS}j$ is a steady velocity of magnetization structure under spin polarized current $Pj$ ($P$ is the spin polarization and $j$ is the applied current density (one-dimensional)).
The spin wave expression is 
\begin{align}
 H_{\rm STT} &= \frac{2\hbar S }{a}v_{\rm st} \lt[ \phi_0+ 2\int dx \lt(  \tilde{\eta}_{\rm R}\nabla_x  \tilde{\eta}_{\rm I}+\frac{1}{\lambda}\tanh u \; \tilde{\eta}_{\rm R}\tilde{\eta}_{\rm I} \rt) \rt]. \label{HSTTsw}
\end{align}
As has been known, a spin-transfer torque contributing to the wall velocity and does not work as a force, as the applied current or $v_{\rm st}$ couples to $\phi_0$ and not to $X$.

The equation of motion for the tree domain wall variables is therefore obtained from Eqs. (\ref{LB1}) (\ref{HSresult}) (\ref{Wresult}) and driving terms (\ref{HBsw})(\ref{HSTTsw}) as 
\begin{align}
 \dot{X}-\alpha\lambda\dot{\phi_0} &= v_{\rm c}\sin2\phi_0 + 2v_{\rm c}\sin2\phi_0 \zeta  +v_{\rm st} \nnr
 \dot{\phi_0}+\alpha\frac{\dot{X}}{\lambda} &= \tilde{B}_z \nnr
 \alpha_\lambda\frac{\dot{\lambda}}{\lambda} &=
   \frac{KS}{\hbar}\lt[\lt(\frac{{\lambda}_0}{\lambda}\rt)^2-(1+\kappa\sin^2\phi_0)\rt]-\dot{\varphi}
     -2\frac{KS}{\hbar}\lt[\lt(\frac{{\lambda}_0}{\lambda}\rt)^2+(1+\kappa\sin^2\phi_0)\rt]\zeta,
     \label{Eqofmo}
\end{align}
where $v_{\rm c}\equiv \frac{KS\kappa}{2\hbar}\lambda$, $\tilde{B}_z\equiv \gamma B_z$ 
and  ${\varphi}$ (Eq. (\ref{varphidef})) and 
\begin{align}
\zeta\equiv \int du \frac{\tanh u}{\cosh u}\tilde{\eta_{\rm R}}, 
\end{align}
are contributions linear in spin wave.

%%%%%%%%%%%%%%%%%%%%%%%%%%%%%%%%%%%%%%%%%%%%%%%%%%%%%%%%%%
\section{Spin wave emission}
%%%%%%%%%%%%%%%%%%%%%%%%%%%%%%%%%%%%%%%%%%%%%%%%%%%%%%%%%%

In this section we study the spin wave emission due to domain wall dynamics. 
The emission is described by the linear coupling between the spin wave field and the domain wall  in Eqs.(\ref{LB1}) (\ref{HSresult}).
Moreover, dynamic second-order couplings in Eqs. (\ref{L2result})(\ref{H2result}) leads to spin wave excitation. 
In the first linear process, the momentum and energy of the spin wave is supplied  by the dynamic domain wall, while the second process
 presents a scattering of spin waves where the domain wall transfer momentum and energy to the incident spin wave. 

%%%%%%%%%%%%%%%%%%%%%%%%%%%
\subsection{Linear emission}
%%%%%%%%%%%%%%%%%%%%%%%%%%%
 
We here discuss the emission due to the linear interactions in Eqs.(\ref{LB1}) (\ref{HSresult}) in the laboratory (rest) frame.
The laboratory frame is described by replacing $\eta(x-X(t),t)$ by  $\eta(x,t)$ in the derivation in Sec. \ref{SECdynamicvariable}.
It turns out that the Lagrangian Eq.(\ref{L2result}) in the laboratory frame has no Doppler shift term and the term 
$\dot{X}\tilde{\eta_{\rm R}} \tilde{\eta_{\rm I}}$ is half. 
The emitted wave has an angular frequency shifted by the Doppler shift from the moving wall. 
Using the equation of motion, Eq. (\ref{Eqofmo}), the spin wave emission arises from the thickness change.
The interaction Hamiltonian reads in the complex notation $\tilde{\eta}=\tilde{\eta}_{\rm R}+i\tilde{\eta}_{\rm I}$
\begin{align}
 H_\eta^{(1)}(t) &= \dot{\lambda}(t)\int dx (\overline{g} \tilde{\eta}+ g\overline{ \tilde{\eta}} ),
 \label{Hsw1}
\end{align}
where 
\begin{align}
 g(x) &\equiv \frac{2\hbar S}{a} \frac{1}{\cosh \frac{x-X(t)}{\lambda}}\lt(-\alpha_\lambda \tanh \frac{x-X(t)}{\lambda}+i\frac{x-X(t)}{\lambda} \rt)
\end{align}
Let us study here the emission treating $\lambda$ as a constant as its dynamics is taken account in the first factor in the interaction Hamiltonian (\ref{Hsw1}).  
The Fourier transform of the interaction is calculated using
\begin{align}
 \int_{-\infty}^\infty du e^{i \tilde{k}u}\frac{u}{\cosh u} &= i\frac{\pi^2}{2} \frac{\sinh \frac{\pi}{2}\tilde{k}}{\cosh ^2\frac{\pi}{2}\tilde{k}} \nnr
 \int_{-\infty}^\infty du e^{i \tilde{k}u}\frac{\tanh u}{\cosh u} &=  \pi  \frac{\tilde{k}}{\cosh \frac{\pi}{2}\tilde{k}} 
\end{align}
as
\begin{align}
 H_\eta^{(1)}(t) &= -\frac{\pi^2}{2} \lambda \dot{\lambda}(t)\sum_{k} \frac{1}{\cosh\frac{\pi}{2}k\lambda}e^{ikX(t)}
 \lt(\tilde{\eta}_{{\rm I}k}(t) \tanh \frac{\pi}{2}k\lambda +\frac{2}{\pi} \alpha_\lambda k\lambda \tilde{\eta}_{{\rm R}k}(t)\rt) ,
 \label{Hsw2}
\end{align}
We consider the case where the wall is approximated by a constant velocity $v_{\rm w}$, i.e., $X(t)=v_{\rm w}t$. 
The frequency representation of time-integral of Eq. (\ref{Hsw2}) is   
\begin{align}
 \int dt H_\eta^{(1)}(t) 
 & = -\frac{\pi^2}{2} \sumOm\sumom \lambda \dot{\lambda}(\Omega)\sum_{k} \frac{1}{\cosh\frac{\pi}{2}k\lambda}
 \lt(\tilde{\eta}_{{\rm I}k}(t) \tanh \frac{\pi}{2}k\lambda+\frac{2\alpha_\lambda}{\pi} k\lambda \tilde{\eta}_{{\rm R}k}(t)\rt) \delta(\omega-(k v_{\rm w}+\Omega)),
 \label{Hsw3}
\end{align}
It is seen that the angular frequency of the emitted spin wave ($\omega$) is $k v_{\rm w}+\Omega$, i.e., that of the thickness variation $\dot{\lambda}$ with a Doppler shift due to the wall motion. 
The Doppler shift of angular frequency, $\delta \nu\equiv k v_{\rm w}$,  is expected to be significant; For $k=1/\lambda$ with $\lambda=10-100$ nm and $v_{\rm w}=100$ m/s, we have 
$\delta\nu=10-1$ GHz.
The function $g(x)$ represents the distribution of the wave vector $k$, which has a broad peak at $k=0$ with a width of the order of $\lambda^{-1}$.

\begin{figure}
 \includegraphics[width=0.6\hsize]{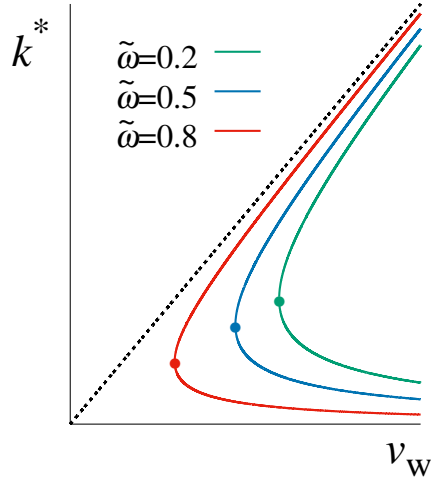}
 \caption{ Plot of the wave length $k^*$ of the emitted spin wave as function of wall velocity $v_{\rm w}$ for $\tilde{\omega}\equiv \hbar\omega_{\phi}/KS=0.2, 0.5$ and $0.8$. Dotted line is $k^*=\frac{\hbar}{KS\lambda^2}v_{\rm w}$. 
 Threshold velocity  for the emission $v_{\rm th}$ is denoted by circles.
 \label{FIGSWemissionk}} 
\end{figure}

To have a finite expectation value $\average{\tilde{\eta}}$, the angular frequency $\omega$ and wave vector $k$ needs to match the dispersion relation of spin wave, $\omega=\omega_k$, i.e.,
\begin{align}
k v_{\rm w}+\hbar \Omega=KS(1+(k\lambda)^2). \label{SWmatching}
\end{align}
The angular frequency $\Omega$ is determined by the equation for $\lambda$ in Eq. (\ref{Eqofmo}), and is of the order of the angular frequency of $\phi_0$, $\omega_\phi$. (See Sec. \ref{SECphirot} for more details.) 
Equation (\ref{SWmatching}) has solution for a velocity larger than the threshold velocity $v_{\rm th}\equiv \frac{2KS}{\hbar}\lambda\sqrt{1-\frac{\hbar\omega_\phi}{KS}}$.
The emitted wave lengths $k^*$ are (plotted in Fig. \ref{FIGSWemissionk})
\begin{align}
 k^* \lambda = \frac{\hbar v_{\rm w}}{2KS\lambda}\lt( 1\pm\sqrt{1-\lt(\frac{v_{\rm th}}{v_{\rm w}}\rt)^2} \rt).
 \label{kstar}
\end{align}
The sign of $k^*$  (direction of emission) is along the wall velocity, meaning that the emission is dominantly in the forward direction.
The group velocity of the emitted wave is of the same order as the wall velocity; 
\begin{align}
\lt. \frac{d\omega_k}{dk}\rt|_{k=k^*}=\frac{2KS}{\hbar} \lambda^2k^*
=v_{\rm w}\lt( 1\pm\sqrt{1-\lt(\frac{v_{\rm th}}{v_{\rm w}}\rt)^2} \rt).
\end{align}

%%%%%%%%%%%%%%%%%%%%%%%%%%
\begin{figure}
 \includegraphics[width=0.4\hsize]{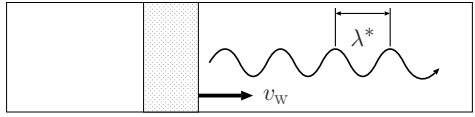}
 \caption{ Schematic figure showing the spin wave emission from a domain wall with thickness oscillation ($\dot{\lambda}$) moving with velocity $v_{\rm w}$.
 The linear coupling leads to a forward emission of spin wave with wave length $\lambda^*\equiv 2\pi/k^*$, where $k^*$ is defined by Eq. (\ref{kstar}). 
 \label{FIGSWemission}} 
\end{figure}
%%%%%%%%%%%%%%%%%%%%%%%%%%%%%%%%%%%%%%
%%%%%%%%%%%%%%%%%%%%%%%%%%%%%%%

%%%%%%%%%%%%%%%%%%%%%%%%%%
%%%%%%%%%%%%%%%%%%%%%%%%%%%%%%%%%%%%%%
%%%%%%%%%%%%%%%%%%%%%%%%%%%%%%%

\begin{figure}
 \includegraphics[width=0.4\hsize]{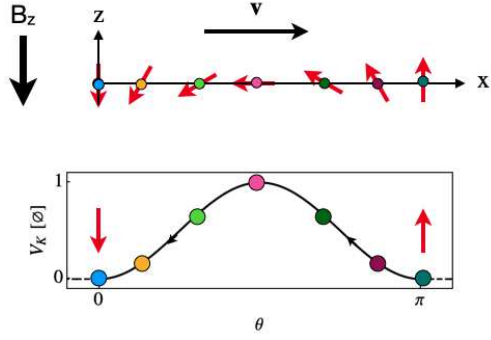}
 \caption{ The local potential $V_K$ for spins in a domain wall arising from the easy axis anisotropy energy, $K$. When the wall moves to the right, the spins right (left) of the wall rotates towards high (low) energy states, resulting in an asymmetric local velocity of rotation, exciting the angle $\phi$ asymmetrically with respect to the wall center. 
 \label{FIGKpot}} 
\end{figure}

The dominant spin wave emission considered here is the antisymmetric excitation of the imaginary part $\tilde{\eta}_{\rm I}$ representing the fluctuation of angle $\phi$. 
The antisymmetric excitation of $\phi$ is a natural excitation arising from the intrinsic property, the anisotropy energy.
The easy-axis anisotropy energy acts as a local potential $V_K$ for each spin in the wall as in Fig. \ref{FIGKpot}.
When the wall moves to the right, the spins ahead of the wall are driven towards the high energy state, while the spins behind (left in Fig. \ref{FIGKpot}) are towards low energy states. 
This asymmetry leads to an asymmetric local ``velocity'' of angle $\theta$, and its  canonical momentum $\phi$. 
This role of $K$ to induce asymmetric $\phi$ is seen in the equations of motion for polar angles \cite{TKS_PR08}:  
Focusing on the contribution of the easy axis anisotropy, the velocity of the in-plane spin rotation, 
$\sin\theta \dot{\phi}=-KS\sin\theta\cos\theta$  is asymmetric with the wall center $\theta=\pi/2$. 
Faster rotation in the left part of the wall ($\frac{\pi}{2}<\theta<\pi$) than the right part ($0<\theta<\frac{\pi}{2}$) indicates that the wall becomes thinner. 
In the equation of motion for $\lambda$  (Eq. (\ref{Eqofmo})), this effect is represented by the term $-\dot{\varphi}$ on the right-hand side, meaning that asymmetric deformation mode $\varphi$ tends to compress the wall.

%%%%%%%%%%%%%%%%%%%%%%%%%%
%%%%%%%%%%%%%%%%%%%%%%%%%%%%%%%%%%%%%%
%%%%%%%%%%%%%%%%%%%%%%%%%%%%%%%

%%%%%%%%%%%%%%%%%%%%%%%%%%%%%%
\subsection{Green's function calculation}
%%%%%%%%%%%%%%%%%%%%%%%%%%%%%%

We present microscopic analysis of the spin wave emission using the Green's function.
We consider here the slow domain wall dynamics limit compared to the spin-wave energy scale and neglect the time-dependence of the variable $u$ arising from variation of $\dot{X}$. The calculation here thus corresponds to the spin wave effects in the  moving frame with the domain wall. 
The amplitude of the spin wave, $\average{\tilde{\eta}}$, is calculated using the path-ordered Green's function method as a linear response to the source field $\dot{\lambda}$.
The amplitude is 
\begin{align}
 \average{\tilde{ \eta}(u,t)} &= -i  \int_{C}dt' \dot{\lambda}(t')\int du'   g(u') 
    \average{T_C \tilde{\eta}(u,t) \overline{\tilde{ \eta}}(u',t') } 
\end{align}
where $C$ denotes the contour for the path-ordered (non-equilibrium) Green's function in the complex time and $T_C$ detnoes the path-ordering.
Evaluating the path-order, we obtain the real-time expression of 
\begin{align}
\average{\tilde{ \eta}(u,t)} &=
\int_{-\infty}^\infty dt' \dot{\lambda}(t')\int du'   g(u') G^\ret_\eta(u,t,u',t')
\end{align}
where 
\begin{align}
  G^\ret_\eta(u,t,u',t')\equiv -i\theta(t-t')  \average{[\tilde{\eta}(u,t), \overline{\tilde{ \eta}}(u',t')] }
\end{align}
  the retarded Green's function of $\tilde{\eta}$. 
The Green's function is calculated expressing $\tilde{\eta}$ in terms of the orthogonal base for spin wave wave function \cite{TKS_PR08} as
\begin{align}
 \tilde{\eta}(u,t)=\sumk\eta_k(t)\phi_k(u),
\end{align}
where  $\phi_k$ is the eigenfunction of Eq. (\ref{SWeigenfunc}) and
$\eta_k$ is the annihilation operator satisfying $[{\eta_k},\overline{\eta}_{k'}]=\delta_{k,k'}$.
The time-development of the operator is $\eta_k(t)=e^{-i\omega_k t}\eta_k(0)$, where $\omega_k\equiv KS\tilde{\omega}_k$ is the energy of spin wave.
The retarded Green's function thus is 
\begin{align}
  G^\ret_\eta(u,t,u',t') = -i\theta(t-t') \sumk  e^{-i\omega_k (t-t')} \phi_k(u) \overline{\phi}_k(u') 
  \equiv \sumom e^{-i\omega(t-t')}  G^\ret_\eta(u,u',\omega)
\end{align}
where 
\begin{align}
G^\ret_\eta(u,u',\omega) &=  \sumk  \frac{1}{\omega-\omega_k+i0} \phi_k(u) \overline{\phi}_k(u')  
\end{align}
is the Fourier transform, $+i0$ denoting the small positive imaginary part.
The Green's function has a nonlocal nature in space, as seen from the overlap of the spin wave function 
\begin{align}
 \sumk \phi_k(u) \overline{\phi}_k(u') &=
  \frac{a}{2\pi \lambda}\lt[\delta(u-u')-\frac{1}{2}
    \lt( e^{-|u-u'|}(1-\tanh u \tanh u')+\sinh (u-u') (\tanh u -\tanh u') \rt) \rt].
    \label{phikoverlap}
 \end{align}
Here we use low-frequency approximation, namely, consider the effect of high-frequency magnon compared to the wall dynamics and use 
$G^\ret_\eta(u,u',\omega) \simeq  -\sumk  \frac{1}{\omega_k} \phi_k(u) \overline{\phi}_k(u')  $.
The retarded Green's function then becomes local in time as 
$ G^\ret_\eta(u,t,u',t') =\delta(t-t') G^\ret_\eta(u,u',\omega)$.
We thus obtain 
\begin{align}
\average{\tilde{ \eta}(u,t)} &= -
 \dot{\lambda}(t)\sumk  \frac{1}{\omega_k} \phi_k(u) \int du'   g(u')\overline{\phi}_k(u')
 \label{etalowfreq}
\end{align}
with $u$ and $u'$ having $X(t)$ of the equal time $t$.
The integral $\int du'   g(u')\overline{\phi}_k(u')$ describing the overlap of spin-wave wave function  and the domain wall is calculated using 
\begin{align}
 \int du \frac{\tanh u}{\cosh u} \overline{\phi}_k(u)
 &= \frac{1}{\sqrt{2\pi\tilde{\omega}_k}}\frac{\pi}{\cosh\frac{\pi}{2}k\lambda} \frac{\tilde{\omega}_k}{2} \nnr
 \int du \frac{u}{\cosh u} \overline{\phi}_k(u)
 &= \frac{1}{\sqrt{2\pi\tilde{\omega}_k}}\frac{\pi}{\cosh\frac{\pi}{2}k\lambda}  
 \label{integrals}
\end{align}
as 
\begin{align}
\int du   g(u)\overline{\phi}_k(u)
 &= \frac{2\hbar S}{a} 
 \frac{1}{\sqrt{2\pi\tilde{\omega}_k}}\frac{\pi}{\cosh\frac{\pi}{2}k\lambda} \lt(i-\alpha_\lambda\frac{\tilde{\omega}_k}{2}\rt)\end{align}
The spin wave amplitude emitted by the wall dynamics is therefore 
\begin{align}
\average{\tilde{ \eta}(u,t)} &= -
 \dot{\lambda}(t)\frac{2\hbar }{K a}\sumk\frac{1}{\sqrt{2\pi\tilde{\omega}_k}}\frac{\pi}{\cosh\frac{\pi}{2}k\lambda}    \frac{1}{\tilde{\omega}_k} \phi_k(u)  \lt(i-\alpha_\lambda\frac{\tilde{\omega}_k}{2}\rt)
 \label{SWamp}
\end{align}
The integral  $\sumk (\tilde{\omega}_k)^{-\beta}\frac{1}{\cosh\frac{\pi}{2}k\lambda}\phi_k(u)$ ($\beta=\frac{1}{2},\frac{3}{2}$) is real, and thus 
$\Re[\average{\tilde{ \eta}}]/\Im[\average{\tilde{ \eta}}]\simeq \alpha$.
As $\average{\tilde{ \eta}(u,t)}$ is odd in $u$, the emitted spin wave is an antisymmetric fluctuation of the angle $\phi$ with respect to the wall center (Fig. \ref{FIGSW}). 
(Because of low frequency approximation in deriving Eq. (\ref{etalowfreq}), the nonlocal nature (Eq. (\ref{phikoverlap})) is smeared out in the result Eq. (\ref{SWamp}). )

The quantities representing the effects of spin wave emission on the wall dynamics in Eq. (\ref{Eqofmo}) are 
\begin{align}
 \zeta &= \int du \frac{\tanh u}{\cosh u} \Re[\tilde{\eta}]
 = \dot{\lambda}  \frac{\pi\hbar }{4Ka} \alpha_\lambda \sumk\frac{1}{\cosh^2\frac{\pi}{2}k\lambda}  
 \equiv  \alpha \mu_\zeta \frac{\dot{\lambda}}{\lambda} \nnr
\varphi &= \int du \frac{ u}{\cosh u} \Im[\tilde{\eta}] 
 = - \dot{\lambda}  \frac{\pi\hbar }{Ka} \sumk \frac{1}{\tilde{\omega}_k^2}\frac{1}{\cosh^2\frac{\pi}{2}k\lambda} 
 \equiv \mu_\varphi \frac{\dot{\lambda}}{\lambda}
 \label{etaamplitudes}
\end{align}
where $\mu_\zeta \equiv \frac{\pi^3\hbar{\lambda} }{48Ka} \sumk\frac{1}{\cosh^2\frac{\pi}{2}k\lambda} $ and
$\mu_\varphi \equiv \frac{\pi\hbar {\lambda}}{Ka} \sumk\frac{1}{\tilde{\omega}_k^2}\frac{1}{\cosh^2\frac{\pi}{2}k\lambda} $.
The first integral is evaluated as 
$\sumk\frac{1}{\cosh^2\frac{\pi}{2}k\lambda} =a\int \frac{dk}{2\pi}\frac{1}{\cosh^2\frac{\pi}{2}k\lambda}=\frac{2a}{\pi^2\lambda}$ and the second one is 
$\sumk\frac{1}{\tilde{\omega}_k^2}\frac{1}{\cosh^2\frac{\pi}{2}k\lambda} \equiv \frac{2a}{\pi^2\lambda}\gamma_\varphi$, where $\gamma_\varphi$ is a constant of the order of unity. 
The constants are therefore 
\begin{align}
\mu_\zeta & =\frac{\pi\hbar }{24K}\nnr
\mu_\varphi & = - \frac{2\hbar \gamma_\varphi}{\pi K} . 
\label{muphi}
\end{align}
From Eq. (\ref{etaamplitudes}), the averaged amplitude of the imaginary part of the emitted spin wave is of the order of 
$ \frac{\hbar \dot{\lambda}}{K\lambda}$ (the real part is a factor of $\alpha$ smaller).
As seen from Eq. (\ref{Eqofmo}), the time scale of $\lambda$ dynamics is $K/\hbar$, and thus the emitted spin wave amplitude can be of the order of unity if the modulation of $\lambda$ is strong, resulting in a significant damping. 
(See Eq. (\ref{alphaunity}) below.)

%%%%%%%%%%%%%%%%%%%%%%%%%%%%%%%%%%%%%%%%%%%%
\subsection{Spin wave excitation due to second order interaction \label{SEC2ndorder}}
%%%%%%%%%%%%%%%%%%%%%%%%%%%%%%%%%%%%%%%%%%%%
Besides emission due to the linear order interaction discussed above, spin waves are excited also  due to the second order interaction in Eqs.(\ref{LB1}) (\ref{HSresult}) when the wall is dynamic.
Here we focus on the effect of a dynamic potential in the Hamiltonian (Eq. (\ref{Hsw})) 
\begin{align}
V(x,t) &\equiv  
4\frac{K S^2 }{a} \frac{1}{\cosh^2 \frac{x-X(t)}{\lambda}}
\label{Vxt}
\end{align}
and calculate the excited spin wave density in the laboratory frame by use of linear response theory.
For a constant wall velocity, $X(t)=v_{\rm w}t$, the Fourier representation of the potential is 
\begin{align}
V_q(\Omega) & =   8\pi^2 \frac{K S^2 \lambda}{a} \frac{q\lambda }{\sinh \frac{\pi}{2}q\lambda} \delta(\Omega-q v_{\rm w}),
\label{Vkomega}
\end{align}
The potential therefore induces Doppler shift of $qv_{\rm w}$
in the angular frequency of the scattered spin wave. 
This dynamic potential induces an excited spin wave density as $ \delta n(x,t) = iG_\eta^<(x,t,x,t) $, where $G_\eta^<$ is the 
lesser Green's function of spin wave. 
The linear response contribution in the Fourier representation is 
\begin{align}
 \delta n(q,\Omega) &=
 i\sum_{k}\sumom V_q(\Omega) (n(\omega+\Omega)-n(\omega))g^\ret_{k\omega}g^\adv_{k+q,\omega+\Omega}
\end{align}
In this process, the excited spin wave density has the same wavelength and angular frequency of the driving potential $V_q(\Omega)$.
This means that the excitation moves together with the domain wall, and thus this is not an emission process.
For slow limit, $q\ll k$ and $\Omega \ll\omega$,  using $n(\omega+\Omega)-n(\omega)=n(\omega+qv_{\rm w})-n(\omega)\simeq qv_{\rm w}n'(\omega)$,   we obtain a compact expression of
\begin{align}
 \delta n(q,\Omega) & = i4\pi \frac{KS^2}{a}v_{\rm w} 
  \frac{(q\lambda)^2  }{\sinh \frac{\pi}{2}q\lambda} \delta(\Omega-qv_{\rm w}) \sumom \sum_k n'(\omega) |g^\ret_{k\omega}|^2 
\end{align}
and the real space profile is 
\begin{align}
 \delta n(x,t)= \delta n_0 \frac{v_{\rm w}}{v_{\rm a}} \frac{\tanh \frac{x-v_{\rm w}t}{\lambda}}{\cosh^2 \frac{x-v_{\rm w}t}{\lambda}}
\end{align}
where $\delta n_0=-\frac{4}{\pi}(KS)^2  \sumom \sum_k n'(\omega) |g^\ret_{k\omega}|^2$ and $v_{\rm a}\equiv K\lambda/\hbar$ is a velocity scale determined by magnetic anisotropy energy.
The induced spin wave density has thus an antisymmetric spatial profile with respect to the wall center and propagate with a domain wall velocity in the present slowly varying limit.
It is not therefore a spin wave emission, but represents the deformation of the wall asymmetric with respect to the center.

%%%%%%%%%%%%%%%%%%%%%%%%%%%%%%%%%%%%%%%%%%%%%%%%%
\section{Equation of motion of three collective coordinates}
%%%%%%%%%%%%%%%%%%%%%%%%%%%%%%%%%%%%%%%%%%%%%%%%%
The equation of motion (\ref{Eqofmo}) including the spin wave emission effects explicitly is therefore
\begin{align}
 \dot{X}-\alpha\lambda\dot{\phi_0} &= v_{\rm c}\sin2\phi_0 + 2v_{\rm c}\sin2\phi_0  \alpha \mu_\zeta \frac{\dot{\lambda}}{\lambda} +v_{\rm st} \label{Xdot1} \\
 \dot{\phi_0}+\alpha\frac{\dot{X}}{\lambda} &=\tilde{B}_z \nnr
 \alpha_\lambda\frac{\dot{\lambda}}{\lambda} &=
   \frac{KS}{\hbar}\lt[\lt(\frac{{\lambda}_0}{\lambda}\rt)^2-(1+\kappa\sin^2\phi_0)\rt]-\mu_\varphi \frac{\ddot{\lambda}}{\lambda}
     -2\frac{KS}{\hbar}\lt[\lt(\frac{{\lambda}_0}{\lambda}\rt)^2+(1+\kappa\sin^2\phi_0)\rt] \alpha \mu_\zeta \frac{\dot{\lambda}}{\lambda}.
     \label{Eqofmosw}
\end{align}
The spin-wave contribution of the first equation, the second term of the right-hand side, is of the order $\alpha$ smaller than the first term and is neglected. 
From the equations, we see that the dynamics of $X$ and $\phi$ are not strongly coupled to the variation of the width.
In particular, when $\kappa$ is small, the dynamics of the wall center ($X$ and $\phi$) governed by the energy scale of $K_\perp=\kappa K$ is much slower than that of a deformation mode $\lambda$, which is of the energy scale of $K$, and thus it is natural that the two  dynamics are decoupled. 
Then $\kappa$ is not small, $\lambda$ affects much the wall center dynamics.

For static case of $\lambda$, we have 
\begin{align}
{\lambda} &=\frac{{\lambda}_0}{\sqrt{1+\kappa\sin^2\phi_0}} ,
     \label{Eqlambda}
\end{align}
as was argued in Refs. \cite{Schryer74,Thiaville04}.
Using this relation assuming slow dynamics to estimate the spin-wave contribution in the equation for $\lambda$, we obtain 
\begin{align}
\mu_\varphi \ddot{\lambda}
+ \tilde{\alpha}_\lambda{\dot{\lambda}} &=
   \frac{KS}{\hbar}{\lambda}\lt[\lt(\frac{{\lambda}_0}{\lambda}\rt)^2-(1+\kappa\sin^2\phi_0)\rt] ,
     \label{Eqofmolambda}
\end{align}
where $\tilde{\alpha}_\lambda \equiv \alpha_\lambda\lt(1+\frac{2S}{\pi}\rt)=\frac{\pi^2}{12}\alpha\lt(1+\frac{2S}{\pi}\rt)$ 
is the effective damping for the width.
The mass for $\lambda$,  $\mu_\varphi$, was induced by the imaginary part of the spin-wave.

%%%%%%%%%%%%%%%%%%%%%%%%%%%%%%%%%%%%%%%%%%%%
\section{Dissipation due to spin wave emission}
%%%%%%%%%%%%%%%%%%%%%%%%%%%%%%%%%%%%%%%%%%%%

Considering the action, which is a time-integral of the Lagrangian, and  by use of integral by parts with respect to time, the linear interaction Hamiltonian, Eq. (\ref{Hsw1}), is equivalent to  
$H_\eta^{(1)} = -{\lambda}F_\lambda$, where 
\begin{align}
F_\lambda\equiv 2\int du \Re[\overline{g} \dot{\tilde{\eta}}],
 \label{Hsw1F}
 \end{align}
is a generalized force for variable $\lambda$.
Using Eqs. (\ref{SWamp})(\ref{integrals}), it reads 
\begin{align}
F_\lambda= - \ddot{\lambda} f_\lambda,
 \label{Hsw1F2}
 \end{align}
where (neglecting the order of $\alpha^2$)
\begin{align}
 f_\lambda \equiv \frac{\pi\hbar^2 S}{Ka^2} \sumk \frac{1}{\tilde{\omega}_k^2}\frac{1}{\cosh^2 \frac{\pi}{2}k\lambda}
 = \frac{2\hbar^2 S}{\pi K \lambda a}\gamma_\varphi .
\end{align}
The energy dissipation rate due to the spin wave emission is therefore 
\begin{align}
 \frac{d{\cal E}_{\rm sw}}{dt}\equiv -\dot{\lambda}F_\lambda =\frac{f_\lambda}{2}\frac{d}{dt}\dot{\lambda}^2,
\end{align}
and thus ${\cal E}_{\rm sw}=\frac{f_\lambda}{2}\dot{\lambda}^2$.
As is seen from Eq. (\ref{Eqofmolambda}), the intrinsic energy scale governing the dynamics of $\lambda$ is $K$, and thus 
the intrinsic scale of $\dot{\lambda}/\lambda$ is of the order of $K/\hbar$.
The energy dissipation by an intrinsic spin-wave emission is estimated roughly as 
${\cal E}_{\rm sw}^{\rm i}  \simeq K\frac{\lambda}{a}$, 
which is the typical spin wave energy multiplied by the number of spin waves excited in the wall.
The quantity $ \frac{d{\cal E}_{\rm sw}^{\rm i}}{dt}$ corresponds to a dissipation function $W_{\rm sw}^{\rm i}$ induced by the intrinsic spin wave emission. 
Considering the intrinsic frequency of $\lambda$ of the order of $K/\hbar$, the Gilbert damping parameter induced by the intrinsic emission is 
\begin{align}
 \alpha_{\rm sw}^{\rm i} \simeq \frac{2a\lambda}{\hbar S}f_\lambda\frac{K}{\hbar}=\frac{4\gamma_\varphi}{\pi}.
 \label{alphaunity}
\end{align}
This value is of the order of unity ($\gamma_\varphi$ is a constant), meaning that spin wave emission from the thickness change is very efficient in dissipating energy from the wall. 
This result may not be surprising if one notices that the intrinsic energy scale of thickness change is that of easy-axis anisotropy energy $K$, which is the energy scale where significant deformation of the wall is induced.

%%%%%%%%%%%%%%%%%%%%%%%%%%%%%%%%%%%%%%%%
\subsection{Modulation of $\lambda$ due to $\phi_0$ dynamics \label{SECphirot}}
%%%%%%%%%%%%%%%%%%%%%%%%%%%%%%%%%%%%%%%%
In most cases, the dynamics of $\lambda$ is driven by the time-dependence of $\phi_0$ as seen in Eq. (\ref{Eqofmolambda}).
Let us consider this case of a forced oscillation. 
We consider by simplyfying $\phi_0$ grows linear with time,   $\phi_0 =\omega_\phi t$, $\omega_\phi$ being a constant.
Linearizing Eq. (\ref{Eqofmolambda}) using $\lambda=\overline{\lambda}+\delta \lambda$, where 
$\overline{\lambda}\equiv \lambda_0/\sqrt{1+\kappa/2}$ is the average thickness, we have an equation of motion of a forced oscillation, 
\begin{align}
\mu_\varphi \ddot{\delta\lambda}
+ \tilde{\alpha}_\lambda{\dot{\delta\lambda}} 
+\mu_\varphi (\Omega_\lambda)^2{\delta\lambda}
&= \frac{KS}{2\hbar}\overline{\lambda} \kappa \cos2\omega_\phi t,
\end{align}
where 
$\Omega_\lambda=\frac{K}{\hbar}\sqrt{\frac{\pi S}{\gamma_\varphi}\lt(1+\frac{\kappa}{2}\rt)}$ is an intrinsic angular frequency of $\delta\lambda$.
The solution having an external angular frequency of $2\omega_\phi$ is
\begin{align}
 \delta\lambda &= \overline{\delta\lambda} \cos (2\omega_\phi t-\varepsilon_\phi),
\end{align}
where 
\begin{align}
 \overline{\delta\lambda}
 &\equiv  \kappa \overline{\lambda} 
 \frac{\frac{\pi S }{4\gamma_\varphi} (K/\hbar)^2}{\sqrt{(\Omega_\lambda^2-4\omega_\phi^2)^2+4(\tilde{\alpha}_\lambda\frac{\omega_\phi}{\mu_{\varphi}})^2 }} 
\end{align}
is the amplitude of the forced oscillation and  $\varepsilon_\phi\equiv \tan ^{-1}\frac{2\tilde{\alpha}_\lambda \frac{\omega_\phi}{\mu_{\varphi}}}{\Omega_\lambda^2-4\omega_\phi^2}$ is a phase shift.
A resonance occurs for $\omega_\phi=\Omega_\lambda/2$. 
The energy dissipation rate for the emission due to forced oscillation induced by dynamics of $\phi$ is 
\begin{align}
 \frac{d{\cal E}_{\rm sw}^{\phi}}{dt}\simeq \frac{\lambda}{a}\lt(\frac{ \overline{\delta\lambda} }{\overline{\lambda}}\rt)^2 
 \frac{\omega_\phi^3}{K}.
\end{align}
The contribution to the Gilbert damping parameter is obtained from the relation  
$ \frac{d{\cal E}_{\rm sw}^{\phi}}{dt}=\alpha_{\rm sw}^{\phi}(\dot{\lambda}/\lambda)^2$ as 
\begin{align}
 \alpha_{\rm sw}^{\phi} \simeq \frac{\lambda}{a}  \frac{\hbar \omega_\phi }{K}.
 \label{alphaphi}
\end{align}

Let us focus on the periodic oscillation of $\phi_0$, realized for large driving forces, namely, for 
$B_z>\alpha\frac{KS\kappa}{2\hbar \gamma}\equiv B_{\rm W}$ ($\gamma B_z>\alpha v_{\rm c}$) for the field-driven case or
$j>\frac{eS^2}{\hbar P}\frac{\lambda}{a}K\kappa\equiv j_{\rm i}$ ($v_{\rm st}>v_{\rm c}$) for the current-driven case 
($B_{\rm W}$ is the Walker's breakdown field and $j_{\rm i}$ is the intrinsic threshold current \cite{TK04}).
The solution of the equation of motion (\ref{Eqofmosw}) then reads
\begin{align}
\phi_0\simeq \omega_\phi t,
\end{align}
where ($j$ is defined in one-dimension to have the unit of A=C/s)
\begin{align}
\omega_\phi \simeq \tilde{B_z}+ \alpha \frac{v_{\rm st}}{\lambda} =\gamma B_z+\frac{aP}{2eS\lambda}\alpha j.
\end{align}
The Gilbert damping constant due to spin wave emission, Eq. (\ref{alphaphi}), thus grows linearly in the driving fields in this oscillation regime.
Using current-induced torque for a pinned domain wall would be straightforward for experimental observation of this behaviour, although the contribution to the Gilbert damping is proportional to $\alpha$ and not large, $\alpha_{\rm sw}^{\phi} \simeq \alpha \frac{\frac{\hbar P}{e}j }{K}$ (for $S\sim1$, $P\sim1$).

%%%%%%%%%%%%%%%%%%%%%%%%%%%%%%%%
\section{Summary}
%%%%%%%%%%%%%%%%%%%%%%%%%%%%%%%%

We studied spin wave emission from a moving domain wall in a ferromagnet by introducing a deformation mode of thickness modulation as a collective coordinate. 
It was shown that the time-derivative of the thickness $\dot{\lambda}$ has a coupling linear in the spin wave field, resulting in an emission, consistent with previous numerical result \cite{Wang12}.
The dominant emitted spin wave is in the forward direction to the moving domain wall and is strongly polarized in the out-of plane direction, i.e., it is a fluctuation of $\phi$.
The dynamics of $\lambda$ is induced by the variation of the angle of the wall plane, $\phi_0$, as has been noted \cite{Schryer74,Thiaville04}.
For a $\phi_0$ with an angular frequency of $\omega_\phi$, the Gilbert damping parameter as a result of spin wave emission is 
$ \alpha_{\rm sw}^{\phi} \simeq \frac{\lambda}{a}  \frac{\hbar \omega_\phi }{K}$, where $K$ is the easy-axis anisotropy energy ($a$ is the lattice constant).

The present study is in the low energy and weak spin wave regime, and treating the higher energy dynamics with strong spin wave emission is an important future subject.

%%%%%%%%%%%%%%%%%%%%%%%%%%%%%%%%%%%%%%%%%%%%%%%%%%%%%%%%%%%%%%%%%%%5
\acknowledgements
GT thanks Y. Nakatani for discussions.
This work was supported by 
%a Grant-in-Aid for Exploratory Research (No.16K13853)  and 
a Grant-in-Aid for Scientific Research (B) (No. 17H02929) from the Japan Society for the Promotion of Science 
and  
a Grant-in-Aid for Scientific Research on Innovative Areas (No.26103006) from The Ministry of Education, Culture, Sports, Science and Technology (MEXT), Japan.

%%%%%%%%%%%%%%
%\bibliography{/home/tatara/References/15,/home/tatara/References/gt,/home/tatara/References/remarks}
%\bibliography{/home/gt/References/15,/home/gt/References/gt,/home/gt/References/remarks}

\end{document}